# Multi-Objective Approach for Optimal Size and Location of DGs in Distribution Systems

Seyed Mohammad Sajjadi Kalajahi, Sina Baghali, Tohid Khalili, *Student Member, IEEE*, Behnam Mohammadi-Ivatloo, *Senior Member, IEEE,* Ali Bidram*, Senior Member, IEEE*

*Abstract*—In the recent years, due to the economic and environmental requirements, the use of distributed generations (DGs) has increased. If DGs have the optimal size and are located at the optimal locations, they are capable of enhancing the voltage profile and reducing the power loss. This paper proposes a new approach to obtain the optimal location and size of DGs. To this end, exchange market algorithm (EMA) is offered to find the optimal size and location of DGs subject to minimizing loss, increasing voltage profile, and improving voltage stability in the distribution systems. The effectiveness of the proposed approach is verified on both 33- and 69-bus IEEE standard systems.

## I. Introduction

THE growing need for energy and rapid growth of electricity consumption will require new solutions in the electric power distribution systems. Due to environmental and economic constraints, expanding the transmission lines and establishing new fossil fuel-based generation units are not viable. To overcome these constraints, utilizing distributed generation (DG) to compensate for load demands is of great importance. DGs could be installed along with the distribution systems in a variety of capacities from a few kW to 50 MW [1]. To meet the required demand, DGs exploit common energy sources such as fossil fuels and renewable energy sources. Some of these sources are fuel cells, internal combustion motors, micro-turbines, gas turbines, wind turbines, photovoltaic, and biomass. Distribution systems can gain many benefits from DGs including increasing the overall efficiency, increasing the reliability of the system, improving the system's security index, enhancing power quality, reducing greenhouse gasses, and relieving the capacity of distribution and transmission systems [2].

The distribution system is accountable for the major part of the power system's overall power loss. Unlike the transmission system, the ratio of line resistance to line reactance is high in the distribution system which causes more voltage sags and power loss. Using DGs in distribution systems will enhance the voltage level and decrease power loss. Installing DGs in wrong locations may result in even higher power loss, so finding the optimal location and size of DGs has a significant effect on the system's performance. There have been numerous efforts to propose the optimal placement and sizing of DGs using different methods and various algorithms. In [3], for solving the optimization problem of DG allocations, an analytical approach has been proposed to decrease the power loss. In [4], the placement of one DG has been implemented based on a power sensitivity index with an analytical approach. Dynamic programming has been used in [5] for DG allocation to decrease power loss and increase the system's voltage level. In [6], dynamic planning of DGs has been considered in an active distribution network. The active distribution system has also been investigated in [7] to maximize DG capacity without causing any technical problems to the system. The mixed-integer linear programming (MILP) method is employed in [8, 9] for installing DGs.

In [10], [11], genetic algorithm have been used for optimally installing DGs by minimizing the power loss and enhancing the voltage level as the objective functions of the problem. This algorithm has also been used in [12]

Tohid Khalili and Ali Bidram are supported by the National Science Foundation EPSCoR Program under Award #OIA-1757207.
Seyed Mohammad Sajjadi Kalajahi and Behnam Mohammadi-Ivatloo are with the Faculty of Electrical and Computer Engineering, University of Tabriz, Tabriz, Iran (e-mails: Mohammad.sajjadi9452@gmail.com, mohammadi@ieee.org). Behnam Mohammadi-Ivatloo is also with Department of energy Technology, Aalborg University, Aalborg, Denmark. Sina Baghali is with the Faculty of Electrical and Computer Engineering, K. N. Toosi University of Technology, Tehran, Iran (e-mail: baghali.sina@email.kntu.ac.ir). Tohid Khalili and Ali Bidram are with the Department of Electrical and Computer Engineering, University of New Mexico, Albuquerque, USA (e-mails: {khalili, bidram}@unm.edu).

with a multi-objective approach considering several load models for DG placements. In [13], network reconfiguration, DG, and fixed/switched capacitor bank placements have been studied simultaneously to achieve a wholesome improvement in the system's behavior in terms of voltage stability, reduction of power loss and the overload's relieving. Optimal allocation of DGs has been studied in [14] by combining genetic algorithm (GA) and simulated annealing-based methods. Fuzzy-based GA for optimal locating and sizing of DGs has been introduced in [15], [16]. In [17], a hybrid method merging intelligent water drops algorithm and GA has been presented to calculate proper sizing and define the optimal place of DGs. Determining the number of DGs and optimal capacity and placement with fuzzy-based multi-objective particle swarm optimization (PSO) algorithm is studied in [18]. The binary PSO-based method is used in [19] for choosing optimal size, place, and time of investment of DGs in the distribution system's planning horizon. In [20], by combining GA and PSO algorithms, the problem of locating and determining the capacity of DGs has been solved with the aim of power loss reduction, voltage stability, and voltage regulation improvement. In [21], Quantum inspired-PSO (Q-PSO) is executed for optimizing the DG placement problem in the power system. Reducing power loss and enhancing voltage stability are the main objectives of the DG placement algorithm in [22]. In [23], chaos embedded symbiotic organism search (CSOS) algorithm has been deployed to minimize the power loss by verifying the optimal location and capacity of DGs. In [24], a multi-objective optimization (MOO) has been introduced to optimize the cost of energy, the power loss, and improve the reliability by optimal sizing and sitting of DGs. Ant-lion optimizer (ALO) is utilized in this paper to optimize the problem. DG placement in [25] is performed utilizing a clonal selection-based artificial immune system method. In [26], optimal placement and sizing of DGs have been studied alongside adjusting the electricity price of DG to maximize DG owners' profit and minimize distribution company cost to meet the demands of the network. A scatter search (SS) method has been used in [26].

In this paper, to solve the problem of optimal placement and sizing and siting of DGs in distribution systems, the exchange market algorithm (EMA) [27] is utilized. EMA is inspired by human behaviors and reactions in trade sharing at stock markets [28]-[33]. The objective of the proposed optimization approach is to minimize power loss, improve voltage profile, and enhance voltage stability in the distribution systems. The effectiveness of the proposed approach is verified on both 33- and 69-bus IEEE standard systems. The remainder of this paper is organized as: Section II presents the problem formulation. The methodology is discussed in Section III. Section IV provides the obtained simulation results. Finally, the relevant conclusions of the optimal placement and sizing of the DGs in the distribution system are presented in Section V.

## II. PROBLEM FORMULATION

The main objectives are minimization of the total power loss and improvement of the voltage profile as well as enhancement of the voltage stability by finding the optimal size and location of DGs. To this end, a multi-objective function is introduced that includes three single objective functions. DGs can provide any amount of power within their minimum and maximum capacity and they can be placed in every bus of the system.

### A. Objective Function

The multi-objective function is as follows [23]:

$$OF = F_1 + P_1 \times F_2 + P_2 \times F_3 \quad (1)$$

where $F_1$, $F_2$, and $F_3$ are three single objective functions. Also, $P_1$ and $P_2$ are penalty factors of the second and third objective functions. In this work, $P_1$ and $P_2$ are equal to 0.6 and 0.35, respectively. The total power loss of the distribution system is introduced as $F_1$ and is presented as follows [20]:

$$F_1 = \sum_{j=1}^{n_b-1} R_j I_j^2 \quad (2)$$

where $I_j$ and $R_j$ show the current and resistance of the $j^{th}$ branch of the system, respectively. The number of buses is $n_b$. $F_2$ defines the total voltage deviation as [23]:

$$F_2 = \sum_{i=1}^{n_b} (V_i - V_{rated})^2 \quad (3)$$

where $V_i$ and $V_{rated}$ are the per-unit voltage of $i^{th}$ bus and rated voltage of system (e.g., 1 pu), respectively.

$F_3$ is selected using the Voltage Stability Index (VSI). Fig.1 shows a distribution system branch between two

buses. The relationship between load's active and reactive power at bus $i$, $P_i$ and $Q_i$, bus $i$ and $j$ voltage, $V_i$ and $V_j$, and line parameters, $X_{ij}$ and $R_{ij}$, expressed as [23]:

$$P_i - jQ_i = V_i^* I_{ij} \quad (4)$$

$$I_{ij} = \frac{V_i - V_j}{R_{ij} + jX_{ij}} \quad (5)$$

Using (4) and (5), one can define VSI as:

$$VSI_i = |V_j|^4 - 4[P_i X_{ij} - Q_i R_{ij}]^2 - 4[P_i R_{ij} + Q_i X_{ij}]|V_j|^2 \quad i=2,3,4,...,n_b \quad (6)$$

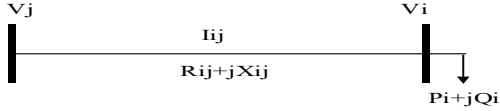

Fig. 1. Diagram of a distribution system's branch.

Normally, the VSI is a positive. By maximizing VSI, voltage stability is improved. So, $F_3$ is defined as:

$$F_3 = \frac{1}{VSI} \quad (7)$$

### B. Constraints

In this paper, the system's inequality and equality constraints are considered. Equality constraints are based on the power flow equations as follows [23]:

$$P_{g,i} - P_i^d - V_i \sum_{j=1}^{Nb} V_j Y_{ij} \cos(\delta_i - \delta_j - \theta_{ij}) = 0 \quad (8)$$

$$Q_{g,i} - Q_i^d - V_i \sum_{j=1}^{Nb} V_j Y_{ij} \sin(\delta_i - \delta_j - \theta_{ij}) = 0 \quad (9)$$

$P_{g,i}$ and $Q_{g,i}$ are generated active and reactive power of $i^{th}$ bus, respectively. Parameters $P_i^d$ and $Q_i^d$ are demanded active and reactive power of $i^{th}$ bus, respectively; $V_i$ and $V_j$ are the voltage of $i^{th}$ and $j^{th}$ bus, respectively; $\delta_i$ and $\delta_j$ are $i^{th}$ and $j^{th}$ bus voltage angle, respectively; $Y_{ij}$ and $\theta_{ij}$ are admittance magnitude and angle.

The inequality constraints in the optimal placement and sizing of DGs' problem are as follows [23]:

$$V_i^{min} \leq V_i \leq V_i^{max} \quad (10)$$

$$I_{ij} \leq I_{ij}^{max} \quad (11)$$

$$P_{i,min}^{DG} \leq P_i^{DG} \leq P_{i,max}^{DG} \quad (12)$$

$$Q_{i,min}^{DG} \leq Q_i^{DG} \leq Q_{i,max}^{DG} \quad (13)$$

Equations (10)-(13) represent the voltage limitations, maximum loading of distribution lines, active power and reactive power limitations of DGs, respectively.

### III. METHODOLOGY

EMA has been inspired by the reaction and decision making of stockholders in the stock market. The two main states of market, unbalanced and balanced modes, are taken into account in every implementation of this algorithm. The significant difference between the two modes is that in balance mode members take action based on superior member's experiences and peruse gaining profit by searching around optimum point. On the other hand, in unbalanced mode, members are prone to take risks in order to get more interests by exploring unknown optimum solutions. In the final stage of the market modes, the dealers, based on their financial state, are categorized into three groups known as groups one, two, and three. EMA approach has faster convergence compared to other algorithms due to utilizing two operators. In a not oscillation condition, EMA utilizes absorptive operators and in the unbalanced mode, the explorer operators are used to find the optimal point [27]. EMA consists of several stages to find the appropriate solution. These steps are explained in Table I.

TABLE I
EMA ALGORITHM

| | |
|---|---|
| 1 | **Input** Data; |
| 2 | **Initialize** Iteration to 1; |
| 3 | Calculate the members' cost by (1) and rank them; |
| 4 | Iteration=Iteration+1; |
| 5 | Change second group members' shares in balanced mode by; |
| 6 | Change third group members' shares in balanced mode; |
| 7 | Analyze the members again and rank them by (1); |
| 8 | Trade second group members' shares in unbalanced mode; |
| 9 | Trade third group members' shares in the unbalanced mode; |
| 10 | **if** Is there termination criterion? **then** |
| |    Save the optimal values; |
| | **else** |
| |    Repeat the steps starting step 3; |
| | **End if** |

In this study, shareholders (members) are DGs' placement and output power (controlling variables). Accordingly, the shares and costs are the value of the controlling variables and the value of the objective function, respectively. After implementing the first iteration, the market reaches a balanced mode, and shares are determined based on the equations mentioned in [27]. The cost function is reevaluated based on new share values and results are classified then the oscillation mode begins. In this step, share values are altered. Afterward, the cost function is calculated with the new share amounts, and classification is done in the same manner.

This loop continues until the algorithm reaches the termination criteria (maximum iteration number).

## IV. SIMULATION RESULTS

The proposed DG placement and sizing approach are implemented on the IEEE 33- and 69-bus radial test systems. The results of EMA are compared with other optimization algorithms to show the effectiveness of the proposed approach. Three DGs are required in the distribution systems and they can provide power in the range of their minimum and maximum values. The maximum and minimum power and power factor of DGs are 1.2 (MVA), 0, and 1, respectively.

### A. Case Study of IEEE 33-Bus Test System

The first test system is the IEEE 33-bus test system, the data of this system are derived from [34]. Each objective function is calculated without and with DGs. Table II shows the objective function values without DGs.

TABLE II
OBJECTIVE FUNCTIONS WITHOUT DGS IN THE 33-BUS TEST SYSTEM

| System | $F_1$ | $F_2$ | $F_3$ | Obj-Function |
|---|---|---|---|---|
| 33-bus | 0.2109 | 0.1338 | 1.4988 | 0.8157 |

Table III presents the results in the presence of DGs (all the numbers are in per-unit with the base values of 1 (MW) and 12.66 (kV)). In this table, different optimization algorithms are used to find the optimal DGs' location and size. While CSOS results in the maximum power loss reduction, EMA has a better solution in minimizing the voltage deviation and maximizing the VSI. EMA has the best answer for the multi-objective function which shows that it can find the optimal size and site of DGs accurately. Fig. 2 shows the voltage profile without and with DGs by EMA, red curve and blue curve represent voltage profile before and after DGs placement, respectively. So, EMA has improved the voltage profile.

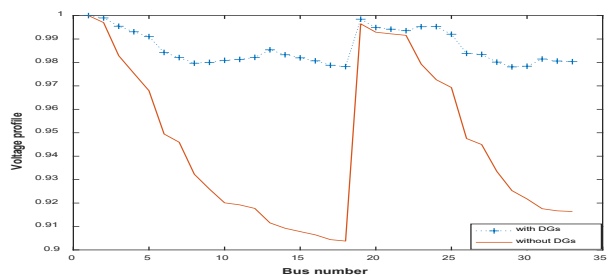

Fig. 2. Voltage profile before and after DG placement in 33-bus system.

### B. Case Study of IEEE 69-Bus Test System

The second case study is the IEEE 69-bus test system. This test system data is extracted from [35]. Three DGs are located on the 69-bus test system with a maximum output of 1.2 (MVA) and the power factor of 1. Table IV shows the objective function values in the system without DGs. Table V summarizes the results in the presence of DGs. In this table, different optimization algorithms are used to find the optimal DGs' location and size. As seen, EMA has a better solution in minimizing the voltage deviation and maximizing the VSI. Moreover, EMA has the best answer for the multi-objective function that shows that EMA can find the optimal size and site of DGs accurately. Fig. 3 shows the voltage profile without and with DGs by EMA. As seen, EMA has improved the voltage profile. EMA has fast convergence and finds the optimal solution with less than 50 iterations.

TABLE IV
OBJECTIVE FUNCTIONS WITHOUT DGS IN THE 69-BUS TEST SYSTEM

| System | $F_1$ | $F_2$ | $F_3$ | Obj-Function |
|---|---|---|---|---|
| 69-bus | 0.225 | 0.0993 | 1.4635 | 0.7968 |

### C. Statistical Analysis of Outputs

This section is dedicated to calculating the mean and variance of the objective function for the IEEE 33- and 69-bus standard systems. The best and worst values obtained from these test systems, with the initial population of fifty, are included in Table VI. The comparison of the results mentioned in this table shows that a small deviation can be observed between the best and worst cases. And, the standard deviation (SD) for objective function is small for both cases and the results presented in Table VI verify the effectiveness of EMA in finding the optimal solution.

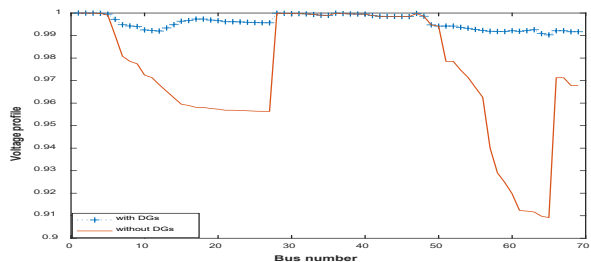

Fig. 3. Voltage profile before and after DG placement in 69-bus system.

TABLE VI
MEAN, VARIANCE, BEST, WORST, AND SD VALUES BY EMA

| System | Mean | Best | Worst | Variance | SD |
|---|---|---|---|---|---|
| 33-bus | 0.4695 | 0.4641 | 0.4707 | $5.41 \times 10^{-6}$ | 0.0023 |
| 69-bus | 0.4451 | 0.4401 | 0.4477 | $1.27 \times 10^{-5}$ | 0.0036 |

TABLE III
OPTIMAL LOCATING AND SIZING OF DGS FOR THE IEEE 33-BUS TEST SYSTEM

| Algorithms | Optimal location | Optimal size (MW) | $F_1$ (pu) | $F_2$ (pu) | VSI (pu) | $F_3$ (pu) | Obj-function (1) | Loss reduction (%) |
|---|---|---|---|---|---|---|---|---|
| GA/PSO [20] | 11 | 0.925 | 0.1034 | 0.0124 | 0.9508 | 1.0517 | 0.4789 | 50.97 |
|  | 16 | 0.863 |  |  |  |  |  |  |
|  | 32 | 1.2 |  |  |  |  |  |  |
| GA [20] | 11 | 1.500 | 0.1063 | 0.0407 | 0.9490 | 1.0537 | 0.4995 | 49.62 |
|  | 29 | 0.423 |  |  |  |  |  |  |
|  | 30 | 1.071 |  |  |  |  |  |  |
| PSO [20] | 8 | 1.177 | 0.1053 | 0.0335 | 0.9256 | 1.0804 | 0.5035 | 50.09 |
|  | 13 | 0.982 |  |  |  |  |  |  |
|  | 32 | 0.829 |  |  |  |  |  |  |
| CSOS [23] | 14 | 0.754 | 0.0714 | 0.0123 | 0.8909 | 1.1224 | 0.3905* | 66.13 |
|  | 24 | 1.099 |  |  |  |  |  |  |
|  | 30 | 1.072 |  |  |  |  |  |  |
| EMA | 13 | 0.963 | 0.077 | 0.0078 | 0.9152 | 1.0925 | 0.4641 | 63.48 |
|  | 24 | 1.168 |  |  |  |  |  |  |
|  | 31 | 1.126 |  |  |  |  |  |  |

*According to (1), the value of objective function calculated by CSOS is not valid and should be 0.4716 instead of 0.3905.

TABLE V
OPTIMAL PLACEMENT AND SIZING OF DGS FOR THE 69-BUS TEST SYSTEM

| Algorithms | Optimal location | Optimal size (MW) | $F_1$ (pu) | $F_2$ (pu) | VSI (pu) | $F_3$ (pu) | Obj-function (1) | Loss reduction (%) |
|---|---|---|---|---|---|---|---|---|
| GA/PSO [20] | 21 | 0.910 | 0.0811 | 0.0031 | 0.9768 | 1.0237 | 0.4413 | 63.90 |
|  | 61 | 1.193 |  |  |  |  |  |  |
|  | 63 | 0.885 |  |  |  |  |  |  |
| GA [20] | 21 | 0.929 | 0.0890 | 0.0012 | 0.9705 | 1.0303 | 0.4503 | 60.39 |
|  | 62 | 1.075 |  |  |  |  |  |  |
|  | 64 | 0.985 |  |  |  |  |  |  |
| PSO [20] | 17 | 0.992 | 0.0832 | 0.0049 | 0.9676 | 1.0334 | 0.4478 | 62.97 |
|  | 61 | 1.199 |  |  |  |  |  |  |
|  | 63 | 0.795 |  |  |  |  |  |  |
| CSOS [23] | 17 | 0.537 | 0.0715 | 0.0063 | 0.9223 | 1.0842 | 0.3981* | 68.16 |
|  | 61 | 1.2 |  |  |  |  |  |  |
|  | 64 | 0.536 |  |  |  |  |  |  |
| EMA | 17 | 0.650 | 0.0755 | 0.0017 | 0.9624 | 1.039 | 0.4401 | 66.44 |
|  | 61 | 1.2 |  |  |  |  |  |  |
|  | 63 | 0.857 |  |  |  |  |  |  |

*According to (1), the value of objective function calculated by CSOS is not valid and should be 0.4547 instead of 0.3981.

## V. CONCLUSION

In this paper, a multi-objective function is proposed to find the optimal size and location of DGs. The main objectives of the optimization problem are minimizing the power loss and voltage deviation and maximizing the VSI. To solve this problem, a metaheuristic algorithm applied which is inspired by the stock market which is called EMA. The EMA has two absorptive and explorer operators which enable it to have fast convergence compared to other algorithms. To prove the effectiveness of EMA, the proposed approach is implemented on two 33- and 69-bus test systems. The results show that EMA obtains a better solution in comparison to CSOS, PSO, GA, and GA/PSO for the multi-objective optimization problem. EMA method takes less than 50 iterations to find the solution of optimal location and size of DGs.

## REFERENCES

[1] T. Ackermann, G. Andersson, and L. Söder, "Distributed generation: a definition," *Electr. Power Syst. Res.*, vol. 57, no. 3, pp. 195–204, Apr. 2001.

[2] B. Singh, V. Mukherjee, and P. Tiwari, "A survey on impact assessment of DG and FACTS controllers in power systems," *Renew. Sustain. Energy Rev.*, vol. 42, pp. 846–882, Feb. 2015.

[3] C. Wang and M. H. Nehrir, "Analytical Approaches for Optimal Placement of Distributed Generation Sources in Power Systems," *IEEE Trans. Power Syst.*, vol. 19, no. 4, pp. 2068–2076, Nov. 2004.

[4] M. M. Aman, G. B. Jasmon, H. Mokhlis, and A. H. A. Bakar, "Optimal placement and sizing of a DG based on a new power stability index and line losses," *Int. J. Electr. Power Energy Syst.*, vol. 43, no. 1, pp.

1296–1304, Dec. 2012.

[5] N. Khalesi, N. Rezaei, and M.-R. Haghifam, "DG allocation with application of dynamic programming for loss reduction and reliability improvement," *Int. J. Electr. Power Energy Syst.*, vol. 33, no. 2, pp. 288–295, Feb. 2011.

[6] K. Zare, S. Abapour, and B. Mohammadi-Ivatloo, "Dynamic planning of distributed generation units in active distribution network," *IET Gener. Transm. Distrib.*, vol. 9, no. 12, pp. 1455–1463, Sep. 2015.

[7] S. Abapour, K. Zare, and B. Mohammadi-Ivatloo, "Maximizing penetration level of distributed generations in active distribution networks," in *2013 Smart Grid Conference (SGC)*, 2013, pp. 113–118.

[8] A. C. Rueda-Medina, J. F. Franco, M. J. Rider, A. Padilha-Feltrin, and R. Romero, "A mixed-integer linear programming approach for optimal type, size and allocation of distributed generation in radial distribution systems," *Electr. Power Syst. Res.*, vol. 97, pp. 133–143, Apr. 2013.

[9] A. Keane and M. O'Malley, "Optimal distributed generation plant mix with novel loss adjustment factors," in *2006 IEEE Power Engineering Society General Meeting*, 2006, p. 6 pp.

[10] C. L. T. Borges and D. M. Falcão, "Optimal distributed generation allocation for reliability, losses, and voltage improvement," *Int. J. Electr. Power Energy Syst.*, vol. 28, no. 6, pp. 413–420, Jul. 2006.

[11] D. H. Popović, J. A. Greatbanks, M. Begović, and A. Pregelj, "Placement of distributed generators and reclosers for distribution network security and reliability," *Int. J. Electr. Power Energy Syst.*, vol. 27, no. 5–6, pp. 398–408, Jun. 2005.

[12] D. Singh, D. Singh, and K. S. Verma, "Multiobjective Optimization for DG Planning With Load Models," *IEEE Trans. Power Syst.*, vol. 24, no. 1, pp. 427–436, Feb. 2009.

[13] D. Esmaeili, K. Zare, B. Mohammadi-Ivatloo, and S. Nojavan, "Simultaneous Optimal Network Reconfiguration, DG and Fixed/Switched Capacitor Banks Placement in Distribution Systems using Dedicated Genetic Algorithm," *Majlesi J. Electr. Eng.*, vol. 9, no.4, p.31, 2015.

[14] M. Gandomkar, M. Vakilian, and M. Ehsan, "A combination of genetic algorithm and simulated annealing for optimal DG allocation in distribution networks," in *Canadian Conference on Electrical and Computer Engineering, 2005.*, 2005, pp. 645–648.

[15] K. Vinothkumar and M. P. Selvan, "Fuzzy Embedded Genetic Algorithm Method for Distributed Generation Planning," *Electr. Power Components Syst.*, vol. 39, no. 4, pp. 346–366, Feb. 2011.

[16] Kyu-Ho Kim, Yu-Jeong Lee, Sang-Bong Rhee, Sang-Kuen Lee, and Seok-Ku You, "Dispersed generator placement using fuzzy-GA in distribution systems," in *IEEE Power Engineering Society Summer Meeting,* 2002, pp. 1148–1153.

[17] M. H. Moradi and M. Abedini, "A novel method for optimal DG units capacity and location in Microgrids," *Int. J. Electr. Power Energy Syst.*, vol. 75, pp. 236–244, Feb. 2016.

[18] S. Ganguly, N. C. Sahoo, and D. Das, "Multi-objective particle swarm optimization based on fuzzy-Pareto-dominance for possibilistic planning of electrical distribution systems incorporating distributed generation," *Fuzzy Sets Syst.*, vol. 213, pp. 47–73, Feb. 2013.

[19] A. Soroudi and M. Afrasiab, "Binary PSO-based dynamic multi-objective model for distributed generation planning under uncertainty," *IET Renew. Power Gener.*, vol. 6, no. 2, p. 67, 2012.

[20] M. H. Moradi and M. Abedini, "A combination of genetic algorithm and particle swarm optimization for optimal DG location and sizing in distribution systems," *Int. J. Electr. Power Energy Syst.*, vol. 34, no. 1, pp. 66–74, Jan. 2012.

[21] M. Nazari-Heris, S. Madadi, M. Pesaran Hajiabbas, and B. Mohammadi-Ivatloo, "Optimal Distributed Generation Allocation Using Quantum Inspired Particle Swarm Optimization," (*Springer*, 2018), pp. 419–432.

[22] Mohamed Imran A and Kowsalya M, "Optimal size and siting of multiple distributed generators in distribution system using bacterial foraging optimization," *Swarm Evol. Comput.*, vol. 15, pp. 58–65, Apr. 2014.

[23] S. Saha and V. Mukherjee, "Optimal placement and sizing of DGs in RDS using chaos embedded SOS algorithm," *IET Gener. Transm. Distrib.*, vol. 10, no. 14, pp. 3671–3680, Nov. 2016.

[24] M. J. Hadidian-Moghaddam, S. Arabi-Nowdeh, M. Bigdeli, and D. Azizian, "A multi-objective optimal sizing and siting of distributed generation using ant lion optimization technique," *Ain Shams Eng. J.*, vol. 9, no. 4, pp. 2101–2109, Dec. 2018.

[25] M. P.S. and S. Hemamalini, "Optimal Siting of Distributed Generators in a Distribution Network using Artificial Immune System," *Int. J. Electr. Comput. Eng.*, vol. 7, no. 2, pp. 641-649, Apr. 2017.

[26] A. Pérez Posada, J. Villegas, and J. López-Lezama, "A Scatter Search Heuristic for the Optimal Location, Sizing and Contract Pricing of Distributed Generation in Electric Distribution Systems," *Energies*, vol. 10, no. 10, p. 1449, Sep. 2017.

[27] N. Ghorbani and E. Babaei, "Exchange market algorithm," *Appl. Soft Comput.*, vol. 19, pp. 177–187, Jun. 2014.

[28] A. Jafari, H. Ganjeh Ganjehlou, T. Khalili, and A. Bidram, "A fair electricity market strategy for energy management and reliability enhancement of islanded multi-microgrids," *Appl. Energy*, vol. 270, p. 115170, Jul. 2020.

[29] A. Jafari, H. Ganjeh Ganjehlou, T. Khalili, B. Mohammadi-Ivatloo, A. Bidram, and P. Siano, "A Two-Loop Hybrid Method for Optimal Placement and Scheduling of Switched Capacitors in Distribution Networks," *IEEE Access*, vol. 8, pp. 38892–38906, 2020.

[30] B. Reimer, T. Khalili, A. Bidram, M. J. Reno, and R. C. Matthews, "Optimal Protection Relay Placement in Microgrids," in *2020 IEEE Kansas Power and Energy Conference (KPEC)*, 2020, pp. 1–6.

[31] T. Khalili, M. T. Hagh, S. G. Zadeh, and S. Maleki, "Optimal reliable and resilient construction of dynamic self-adequate multi-microgrids under large-scale events," *IET Renew. Power Gener.*, vol. 13, no. 10, pp. 1750–1760, Jul. 2019.

[32] T. Khalili, A. Jafari, and E. Babaei, "Scheduling and siting of storages considering power peak shaving and loss reduction by exchange market algorithm," in *2017 Smart Grid Conference (SGC)*, 2017, vol. 2018-Janua, pp. 1–7.

[33] T. Khalili, A. Bidram, and M. J. Reno, "Impact study of demand response program on the resilience of dynamic clustered distribution systems," *IET Gener. Transm. Distrib.*, Apr. 2020.

[34] M. E. Baran and F. F. Wu, "Network reconfiguration in distribution systems for loss reduction and load balancing," *IEEE Trans. Power Deliv.*, vol. 4, no. 2, pp. 1401–1407, Apr. 1989.

[35] D. Das, "Optimal placement of capacitors in radial distribution system using a Fuzzy-GA method," *Int. J. Electr. Power Energy Syst.*, vol. 30, no. 6–7, pp. 361–367, Jul. 2008.